\documentclass{article}

\usepackage{arxiv}

\usepackage[utf8]{inputenc}
\usepackage[T1]{fontenc}
\usepackage{lmodern}
\usepackage{amsmath}
\usepackage{amssymb}
\usepackage{graphicx}
\usepackage{booktabs}
\usepackage{array}
\usepackage{xcolor}
\usepackage{listings}
\usepackage{pgfplots}
\pgfplotsset{compat=1.17}
\usepgfplotslibrary{groupplots}
\usetikzlibrary{arrows.meta,positioning,fit,backgrounds}
\usepackage[numbers]{natbib}
\usepackage[hidelinks]{hyperref}

\definecolor{codebg}{rgb}{0.96,0.96,0.96}
\definecolor{codecomment}{rgb}{0.30,0.45,0.30}
\definecolor{coderule}{rgb}{0.80,0.80,0.80}
\definecolor{accent}{rgb}{0.13,0.29,0.53}
\definecolor{accentb}{rgb}{0.80,0.33,0.13}

\lstdefinestyle{logstyle}{
  backgroundcolor=\color{codebg},
  basicstyle=\ttfamily\footnotesize,
  commentstyle=\color{codecomment},
  breaklines=true,
  breakatwhitespace=false,
  captionpos=b,
  frame=single,
  rulecolor=\color{coderule},
  showstringspaces=false,
  keepspaces=true,
  columns=fullflexible,
  xleftmargin=0.5em,
  xrightmargin=0.5em,
  aboveskip=1em,
  belowskip=1em,
}
\newcommand{\code}[1]{\texttt{#1}}

\title{On the Limitations of Non-GPU AI Accelerators for Large-Model
Inference: A Field Study of MoE and Multimodal Serving on Huawei Ascend}

\author{%
  Zheng Yu\thanks{Correspondence: \texttt{zheng.yu3@mail.mcgill.ca}.} \\
  School of Data Science \\
  The Chinese University of Hong Kong, Shenzhen \\
  Shenzhen, Guangdong, China \\
}

\date{\today}

\begin{document}
\maketitle

\begin{abstract}
Non-GPU AI accelerators are increasingly adopted as alternatives to
general-purpose GPUs for large-model inference, motivated by supply, cost, and
availability considerations. Yet the practical cost of migrating a demanding
workload from the mature CUDA ecosystem to such an accelerator is poorly
documented. We report a field study of deploying two concrete, demanding
inference workloads on a 16-device Huawei Ascend~910 system using the CANN
software stack and the \emph{vLLM-Ascend} inference engine. The first is an
\emph{LLM-as-a-judge} value-alignment and safety evaluation pipeline: a
weight-quantized (W8A8) Mixture-of-Experts (MoE) judge model,
DeepSeek-V4-Flash ($\sim$300\,GB), scores the responses of twenty frontier LLMs
(e.g., GPT-5.1, Gemini~3~Pro, Claude Haiku~4.5, DeepSeek-V3.2, Qwen3-Max,
GLM-4.6) at a scale of tens of thousands of prompts each. The second is a
\emph{multimodal medical vision--language} benchmarking workload:
DeepSeek-V4-Flash-Vision ($\sim$540\,GB in \code{bf16}), a frozen MoE language
decoder fused with a Qwen3.5 vision tower through a trained
merger/bridge, evaluated on the MMMU and MMMU-Pro benchmarks. Bringing both
workloads to a serviceable state required twelve source-level patches to the
vendor inference plugin, the deliberate disabling of several throughput features
to preserve numerical correctness, and operational scaffolding to absorb
recurring low-level device faults. We organize our observations into eight
classes of platform-level limitation --- software-stack maturity and
operator/feature-coverage gaps, fragile multi-axis parallelism, kernel-level
(aicore/vector-core) numerical faults, immature graph compilation, unstable
advanced features, performance and scalability ceilings, weak operational
observability, and ecosystem fragmentation --- and, for each, give the symptom,
evidence, and likely root cause. We quantify the integration effort, the observed
concurrency behavior, and the end-to-end benchmark quality that confirms both
workloads were served correctly, and we distill a set of general, vendor-agnostic
strategies for adopting alternative accelerators. Our aim is a reproducible
reference for teams evaluating or operating this class of accelerator.
\end{abstract}


\newpage
\tableofcontents
\bigskip
\newpage
\section{Introduction}
The CUDA ecosystem's dominance in large-model inference has made alternative
accelerators attractive wherever GPU supply, cost, or procurement constraints
apply. Huawei's Ascend line, together with the Compute Architecture for Neural
Networks (CANN) software stack~\cite{cann}, is among the most mature such
alternatives; its DaVinci-architecture NPUs~\cite{davinci,ascendhpca} target the
same transformer~\cite{transformer} inference workloads that GPUs serve, and
open-source engines such as vLLM~\cite{vllm} now have a vendor-maintained Ascend
port~\cite{vllmascend}. On paper, a 16-card Ascend~910 system offers ample memory
and interconnect to host frontier-scale MoE models~\cite{deepseekv3,mixtral} and
multimodal models~\cite{qwen2vl,llava}.

In practice, migrating a non-trivial workload from a GPU/CUDA baseline to this
stack is far from a recompile. Over two independent deployments --- an
LLM-as-a-judge value-alignment evaluation service and a multimodal medical
vision--language benchmarking pipeline (both detailed in
Section~\ref{sec:setup}) --- we accumulated a substantial list of failures,
workarounds, and hard limits that were \emph{not} specific to our models but were
instead properties of the accelerator, its compiler and operator library, and the
vendor inference plugin. These observations are difficult to find in vendor
documentation and costly to rediscover, which motivates this field study.

\paragraph{Contributions.}
\begin{itemize}
  \item A structured catalog of eight classes of platform-level limitation
        (Section~\ref{sec:limitations}) observed while deploying large MoE and
        multimodal inference on Ascend~910 via vLLM-Ascend, each presented as
        \emph{symptom $\rightarrow$ evidence $\rightarrow$ root cause}.
  \item Platform-level evidence --- device fault codes, kernel exception strings,
        engine log excerpts --- and quantitative characterizations of integration
        effort, concurrency behavior, and end-to-end benchmark quality
        (Figures~\ref{fig:arch}--\ref{fig:memory}).
  \item A set of general, vendor-agnostic strategies
        (Section~\ref{sec:mitigations}) for adopting alternative accelerators,
        spanning correctness assurance, capacity planning, fault tolerance, and
        ecosystem engagement.
\end{itemize}

\paragraph{Scope and non-goals.} This is a qualitative field study with
supporting operational and benchmark measurements. Our focus is the
\emph{hardware/software platform}, not the two applications: we describe the
workloads and report end-to-end quality only insofar as they establish that the
deployments were real and correctly served, and we do not claim our patches are
canonical fixes. The limitations we catalog are properties of the accelerator,
its compiler and operator library, and the vendor inference plugin, and we expect
them to generalize beyond the specific models we ran.

\section{Related Work}
\label{sec:related}
\paragraph{Inference serving systems.} High-throughput LLM serving has been
advanced by iteration-level scheduling and selective batching in
Orca~\cite{orca}, and by paged key--value memory management and continuous
batching in vLLM~\cite{vllm}. Efficient attention kernels such as
FlashAttention~\cite{flashattention} and FlashAttention-2~\cite{flashattention2}
underpin the throughput of these systems. Our study concerns what happens when
such a serving stack is ported to a non-GPU accelerator.

\paragraph{Parallelism for large models.} Scaling inference beyond a single
device relies on tensor parallelism~\cite{megatron}, pipeline
parallelism~\cite{gpipe}, and expert/sharded parallelism~\cite{gshard}. We use
these decompositions directly; several of our limitations
(Section~\ref{sec:parallelism}) arise precisely at their interactions on the
target stack.

\paragraph{Mixture-of-experts.} Sparsely-activated MoE
layers~\cite{shazeer,switch} enable large parameter counts at fixed compute and
are used by recent open models~\cite{mixtral,deepseekv3,deepseekv2}. MoE routing
and expert parallelism are among the least-covered paths in the vendor plugin.

\paragraph{Speculative decoding.} Draft-and-verify decoding~\cite{speculative}
and its learned-head variants~\cite{medusa,eagle} accelerate autoregressive
generation. On the target stack the multi-token-prediction variant was the least
stable feature we tried (Section~\ref{sec:advanced}).

\paragraph{Quantization.} Post-training quantization methods including
SmoothQuant~\cite{smoothquant}, GPTQ~\cite{gptq}, and AWQ~\cite{awq} reduce the
memory and bandwidth cost of inference; our MoE workload is served in a W8A8
configuration.

\paragraph{Vision--language models.} Contrastive pretraining~\cite{clip} and
visual instruction tuning~\cite{llava,qwen2vl} produce the vision-tower + LLM
decoder design used by our multimodal workload, whose fusion path stresses the
plugin's multimodal support.

\paragraph{Accelerator architectures.} The DaVinci
architecture~\cite{davinci} and the Ascend NPU~\cite{ascendhpca} provide the
cube/vector compute model targeted here; PyTorch~\cite{pytorch} is the framework
integrated via a device backend. To our knowledge, few public field studies
document the systems-level limitations of serving frontier-scale MoE and
multimodal models on this class of accelerator.

\section{Background}
\label{sec:background}

\subsection{Ascend DaVinci, CANN, HCCL, and torch\_npu}
Ascend NPUs implement Huawei's \emph{DaVinci} architecture~\cite{davinci}, whose
compute core (``aicore'') combines a matrix/\emph{cube} unit for dense tensor
contractions with a \emph{vector} unit for elementwise and reduction operations,
plus a memory-transfer engine (MTE) that moves data across the on-device memory
hierarchy~\cite{ascendhpca}. Software access is mediated by CANN~\cite{cann},
which provides the runtime (ACL), the operator library, a graph compiler, and the
collective-communication library HCCL (the analogue of NCCL). PyTorch
integration is provided by \code{torch\_npu}, a device backend that maps ATen
operators~\cite{pytorch} onto CANN kernels.

\subsection{vLLM-Ascend as a vendor plugin}
vLLM~\cite{vllm} implements high-throughput LLM serving via paged attention and
continuous batching. On Ascend, serving is provided by \emph{vLLM-Ascend}, a
platform plugin that registers an ``ascend'' backend and supplies NPU-specific
model implementations, attention kernels, quantization paths, and a graph-capture
integration. Crucially, it is a separate, fast-moving codebase that reimplements
or overrides substantial portions of upstream vLLM; feature parity with the CUDA
path is partial and version-dependent.

\subsection{Contrast with the CUDA ecosystem}
Two structural differences drive most of what follows. First, graph capture on
Ascend uses ACL Graph rather than CUDA Graph, and there is no TorchInductor-style
just-in-time kernel compiler in the inference path --- kernels come from the
operator library or are compiled ahead of time, so uncovered shapes/operators
fall back to eager execution. Second, the operator and feature coverage of the
vendor stack trails the CUDA reference implementation, so advanced parallelism,
fused kernels, and speculative decoding are newer, more fragile, and more
frequently gated behind experimental flags.

\section{Deployment Setup and Methodology}
\label{sec:setup}
All experience reported here is from a single node with 16 Ascend~910
accelerators running the CANN toolkit inside the vendor-provided vLLM-Ascend
container (engine version in the 0.13 series). We deployed two workloads,
summarized in Table~\ref{tab:workloads} and described in detail below. Both are
built on the DeepSeek-V4-Flash family, whose decoder combines an MoE feed-forward
stack with a DeepSeek Sparse Attention (DSA) path, a heterogeneous-computing
multi-head (MHC) representation, and a Multi-Token-Prediction (MTP) draft head;
these are exactly the components whose Ascend support proved least mature.

\begin{table}[t]
\centering
\caption{The two inference workloads. Both share the DeepSeek-V4-Flash decoder
family; they differ in quantization, modality, and parallel layout.}
\label{tab:workloads}
\small
\begin{tabular}{@{}p{0.17\linewidth}p{0.36\linewidth}p{0.36\linewidth}@{}}
\toprule
 & \textbf{Workload A: value-alignment judge} & \textbf{Workload B: multimodal VLM} \\
\midrule
Task & LLM-as-a-judge value/safety scoring & Vision--language benchmarking (MMMU, MMMU-Pro) \\
Model & DeepSeek-V4-Flash-w8a8-mtp & DeepSeek-V4-Flash-Vision (Qwen3.5 vision tower + MoE decoder) \\
Precision / size & W8A8 quantized, $\sim$300\,GB & \code{bf16}, $\sim$540\,GB \\
Parallelism & TP\,$=$\,8 $\times$ DP\,$=$\,2, expert parallel & TP\,$=$\,8 $\times$ PP\,$=$\,2, expert parallel \\
Graph mode & Full-decode ACL graph capture & \code{enforce-eager} (no capture) \\
Advanced features & MTP speculative decoding (1 token) & --- \\
Batching & \code{max-num-seqs}\,$=$\,128 & \code{max-num-seqs}\,$=$\,8 \\
\bottomrule
\end{tabular}
\end{table}

\subsection{Case study A: an LLM-as-a-judge value-alignment service}
\label{sec:workloadA}
The first workload is an offline batch-inference pipeline that uses a large
language model as an evaluator (``LLM-as-a-judge''). The judge,
DeepSeek-V4-Flash-w8a8-mtp, is served through an OpenAI-compatible endpoint and
scores the responses of twenty contemporary LLMs --- including GPT-5.1,
Gemini~3~Pro, Claude Haiku~4.5, DeepSeek-V3.2, Qwen3-Max, GLM-4.6,
Kimi-K2, Doubao, ERNIE, Hunyuan, Mistral-Large, Llama-4-Scout, and
others --- on questions that probe value-alignment and safety. For each
(question, response) pair the judge applies a weighted rubric (per-question
criteria plus general deduction items) and emits a structured JSON verdict with a
final score. The campaign spans tens of thousands of prompts per model, which
makes \emph{sustained throughput and reliability}, rather than latency, the
binding constraint; this workload is the source of the concurrency and
reliability findings in Sections~\ref{sec:perf}--\ref{sec:reliability}.

\subsection{Case study B: a multimodal medical vision--language model}
\label{sec:workloadB}
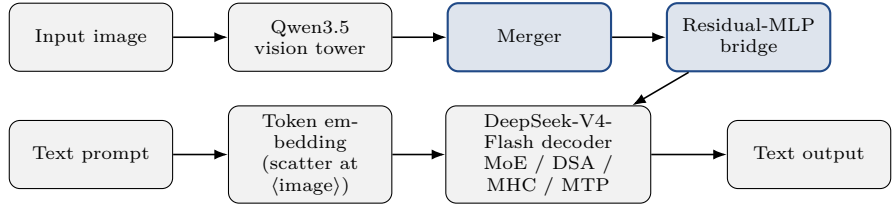
\begin{figure}[t]
\centering
\begin{tikzpicture}[
  box/.style={draw, rounded corners, align=center, font=\scriptsize,
              minimum height=0.9cm, text width=1.95cm, inner sep=3pt},
  trained/.style={box, fill=accent!15, draw=accent, thick},
  frozen/.style={box, fill=black!5},
  arr/.style={-{Latex[length=1.8mm]}, semithick},
]
\node[frozen]  (img) at (0,1.55)  {Input image};
\node[frozen]  (vit) at (2.9,1.55){Qwen3.5\\vision tower};
\node[trained] (mg)  at (5.8,1.55){Merger};
\node[trained] (br)  at (8.7,1.55){Residual-MLP\\bridge};
\node[frozen]  (txt) at (0,0)   {Text prompt};
\node[frozen]  (emb) at (2.9,0) {Token embedding\\(scatter at $\langle$image$\rangle$)};
\node[frozen, text width=2.5cm] (dec) at (6.05,0) {DeepSeek-V4-Flash decoder\\MoE / DSA / MHC / MTP};
\node[frozen]  (out) at (9.5,0) {Text output};
\draw[arr] (img)--(vit);
\draw[arr] (vit)--(mg);
\draw[arr] (mg)--(br);
\draw[arr] (txt)--(emb);
\draw[arr] (br) -- (dec);
\draw[arr] (emb)--(dec);
\draw[arr] (dec)--(out);
\end{tikzpicture}
\caption{Data flow of the multimodal Workload~B. A frozen Qwen3.5 vision tower and
a frozen DeepSeek-V4-Flash MoE decoder are joined by a \emph{trained} merger and
residual-MLP bridge (shaded). Visual features are scattered into the decoder's
token-embedding sequence at image-placeholder positions. The decoder's MoE, DSA
(sparse attention), MHC (heterogeneous-computing heads), and MTP (draft head)
blocks are precisely the paths whose Ascend support was least mature
(Section~\ref{sec:limitations}).}
\label{fig:arch}
\end{figure}

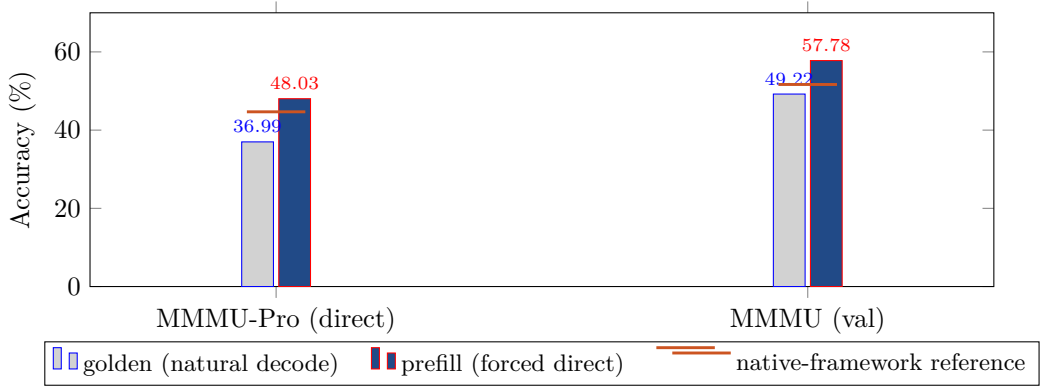
\begin{figure}[t]
\centering
\begin{tikzpicture}
\begin{axis}[
    width=0.82\linewidth, height=5.2cm,
    ybar,
    bar width=0.42cm,
    enlarge x limits=0.35,
    ymin=0, ymax=70,
    ylabel={Accuracy (\%)},
    symbolic x coords={MMMU-Pro (direct), MMMU (val)},
    xtick=data,
    nodes near coords, nodes near coords style={font=\scriptsize},
    legend style={at={(0.5,-0.20)}, anchor=north, legend columns=-1,
                  font=\footnotesize, /tikz/every even column/.append style={column sep=0.35cm}},
]
\addplot+[fill=gray!35] coordinates {(MMMU-Pro (direct),36.99) (MMMU (val),49.22)};
\addplot+[fill=accent]  coordinates {(MMMU-Pro (direct),48.03) (MMMU (val),57.78)};
\addplot+[only marks, mark=-, mark size=11pt, line width=1.1pt, color=accentb,
          every node near coord/.style={opacity=0}]
          coordinates {(MMMU-Pro (direct),44.68) (MMMU (val),51.67)};
\legend{golden (natural decode), prefill (forced direct), native-framework reference}
\end{axis}
\end{tikzpicture}
\caption{End-to-end quality of the multimodal Workload~B (fully instruction-tuned
checkpoint) on MMMU-Pro (\code{standard\_4/direct}, 1730 items) and MMMU
validation (900 items), served on Ascend via vLLM-Ascend. Under the forced-direct
(\emph{prefill}) protocol the port matches or exceeds the native-framework
reference (orange ticks), confirming numerical correctness end-to-end. The
golden/prefill gap reflects a \emph{model} behavior (a reasoning-oriented model
truncated by a 4-token answer budget), not an accelerator defect.}
\label{fig:bench}
\end{figure}

\begin{figure}[t]
\centering
\begin{tikzpicture}
\begin{axis}[
    width=0.9\linewidth, height=3.6cm,
    xbar stacked,
    xmin=0, xmax=64,
    xlabel={Per-card memory (GB)},
    symbolic y coords={Ascend 910 card},
    ytick=data,
    bar width=0.6cm,
    legend style={at={(0.5,-0.55)}, anchor=north, legend columns=-1,
                  font=\footnotesize, /tikz/every even column/.append style={column sep=0.4cm}},
    xtick={0,16,32,48,64},
    enlarge y limits=0.6,
    nodes near coords style={font=\scriptsize},
]
\addplot+[fill=accent] coordinates {(34,{Ascend 910 card})};
\addplot+[fill=accentb] coordinates {(2,{Ascend 910 card})};
\addplot+[fill=gray!35] coordinates {(22.9,{Ascend 910 card})};
\addplot+[fill=gray!12] coordinates {(5.1,{Ascend 910 card})};
\legend{LM decoder shard, Vision tower, KV cache + activations, Reserved / headroom}
\end{axis}
\end{tikzpicture}
\caption{Per-card memory footprint for the $\sim$540\,GB \code{bf16} multimodal
workload split across 16 Ascend~910 devices (TP\,$=$\,8\,$\times$\,PP\,$=$\,2).
Each card holds roughly a 34\,GB decoder shard plus a $\sim$2\,GB replicated
vision tower, with the remainder of the $\sim$59\,GB working budget
(0.92 of 64\,GB) going to KV cache and activations. The large weight footprint
leaves little headroom, which interacts with the tensor-parallel cap discussed in
Section~\ref{sec:parallelism}.}
\label{fig:memory}
\end{figure}
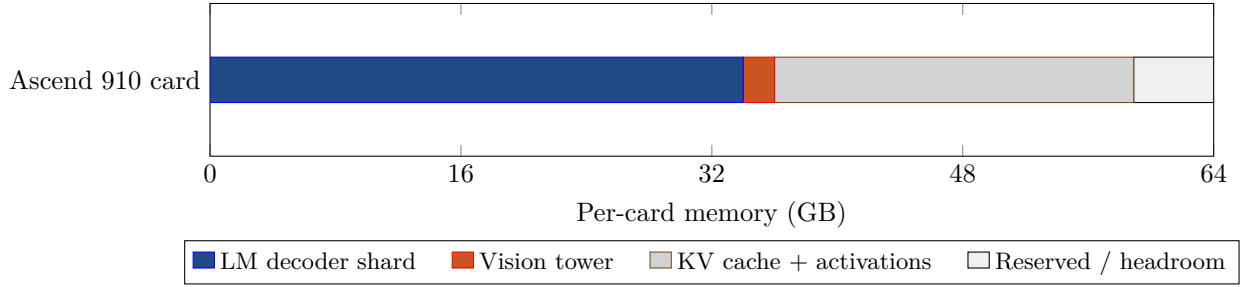

The second workload serves DeepSeek-V4-Flash-Vision, a multimodal model that
fuses a frozen Qwen3.5 vision tower and a frozen DeepSeek-V4-Flash MoE decoder
through a trained merger and a residual-MLP bridge
(Figure~\ref{fig:arch}). Image patches are encoded by the vision tower, projected
by the merger, adapted by the bridge, and scattered into the decoder's token
embedding at the image-placeholder positions; the decoder then generates text
autoregressively. Because the \code{bf16} weights are large
(Figure~\ref{fig:memory}) and an attention head-grouping factor caps tensor
parallelism at eight, the single replica must be split with an additional
pipeline-parallel stage (TP\,$=$\,8\,$\times$\,PP\,$=$\,2), which exercises
pipeline-parallel code paths that the text-only reference deployment never used.
We validated the deployment on the MMMU and MMMU-Pro benchmarks
(Figure~\ref{fig:bench}): under a forced-direct-answer (\emph{prefill}) protocol
the fully instruction-tuned checkpoint reaches $48.0\%$ on MMMU-Pro and $57.8\%$
on MMMU (validation), matching or exceeding the native-framework reference and
confirming that the port is numerically correct end-to-end.

\paragraph{Methodology.} Our evidence is observational, gathered while bringing
the two workloads into production service rather than in a controlled benchmark.
Concretely, we draw on (i) the set of source patches required to initialize and
correct the plugin, (ii) engine and device logs captured during startup and
sustained serving, (iii) coarse operational measurements (request concurrency
versus throughput and queue depth, cold-start time, per-card memory), and
(iv) end-to-end benchmark scores used solely to confirm correct operation. We
foreground platform-level artifacts --- vendor/stack component names, environment
variables, device fault codes, and generic exception strings --- because these
are the properties that transfer across deployments. The two workloads exercise
the harder corners of the stack --- expert-parallel MoE routing, the DSA
sparse-attention path, and (for Workload~B) a multimodal input pipeline combined
with pipeline parallelism --- and, as we show next, essentially every one of
those corners required intervention.

Figure~\ref{fig:memory} shows the per-card memory budget for the multimodal
workload, which explains why the tensor-parallel cap forces a second parallel
axis (Section~\ref{sec:parallelism}).

\section{Limitations}
\label{sec:limitations}
We group our observations into eight categories, each presented as
\emph{symptom}, \emph{evidence}, and \emph{root cause}. Table~\ref{tab:summary}
summarizes the categories and the mitigations discussed in
Section~\ref{sec:mitigations}.

\begin{table}[t]
\centering
\caption{Summary of observed limitation classes and mitigations.}
\label{tab:summary}
\small
\begin{tabular}{@{}p{0.05\linewidth}p{0.34\linewidth}p{0.52\linewidth}@{}}
\toprule
\textbf{\S} & \textbf{Limitation class} & \textbf{Practical mitigation} \\
\midrule
5.1 & Stack maturity / operator \& feature coverage & Patch the plugin at startup; keep a versioned patch set \\
5.2 & Fragile multi-axis parallelism (PP$>$1, SP, MC2) & Disable SP/fused-MC2; correctness-first config; accept TP$=$8 cap \\
5.3 & Kernel numerical / stability faults (aicore, MTE) & Watchdog + auto-restart; reduce concurrency; avoid fragile ops \\
5.4 & Immature graph compilation & Prefer \code{enforce-eager}; limit full-decode capture sizes \\
5.5 & Advanced-feature gaps (speculative decoding) & Disable MTP when unstable; trade throughput for reliability \\
5.6 & Performance / scalability ceilings & Tune concurrency to a ``sweet spot''; budget slow startup \\
5.7 & Operational reliability / observability & External health checks; log scraping for device faults \\
5.8 & Ecosystem fragmentation / portability tax & Pin versions; document env-var matrix; patch-on-startup \\
\bottomrule
\end{tabular}
\end{table}

\subsection{Software-stack maturity and operator/feature coverage gaps}
\label{sec:maturity}
\paragraph{Symptom.} Bringing either workload up required a sequence of
source-level patches to the vendor inference plugin before it would even
initialize, let alone produce correct output. In total, twelve distinct patches
were applied programmatically to the installed plugin package at startup.

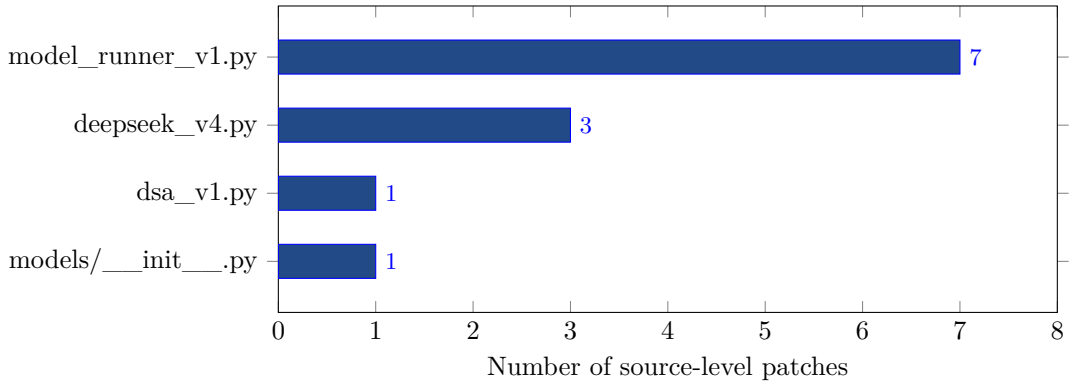
\begin{figure}[t]
\centering
\begin{tikzpicture}
\begin{axis}[
    width=0.72\linewidth, height=4.8cm,
    xbar,
    xmin=0, xmax=8,
    xlabel={Number of source-level patches},
    symbolic y coords={{models/\_\_init\_\_.py},{dsa\_v1.py},{deepseek\_v4.py},{model\_runner\_v1.py}},
    ytick=data,
    nodes near coords, nodes near coords align={horizontal},
    nodes near coords style={font=\footnotesize},
    bar width=0.45cm,
    enlarge y limits=0.25,
    y=0.9cm,
]
\addplot+[fill=accent] coordinates {
    (7,{model\_runner\_v1.py})
    (3,{deepseek\_v4.py})
    (1,{dsa\_v1.py})
    (1,{models/\_\_init\_\_.py})
};
\end{axis}
\end{tikzpicture}
\caption{The twelve startup patches applied to the vendor inference plugin,
grouped by source file. The worker runtime (\code{model\_runner\_v1.py}) and the
model definition (\code{deepseek\_v4.py}) absorb most of the changes, reflecting
the pipeline-parallel, multimodal, and MoE/sparse-attention paths that fell
outside the plugin's tested envelope.}
\label{fig:patches}
\end{figure}

\paragraph{Evidence.} Figure~\ref{fig:patches} breaks the twelve patches down by
plugin source file. They spanned loader, worker, and attention code and addressed
independent defects, including: a pipeline-parallel guard for a per-layer
attention ``sink'' weight (a direct dictionary lookup that raised \code{KeyError}
on ranks that did not own the layer); the memory-profiling dummy run taking an
inconsistent embedding path for hash-routed MoE; the sparse-attention key/value
state and Hadamard transform being gated on a text-only \code{model\_type} and
therefore never allocated for the multimodal config (so the \emph{first real}
prefill --- not startup --- crashed with an out-of-bounds index); the multimodal
embedding merge running on every pipeline-parallel rank rather than only the
first; and a persistent pipeline intermediate-tensor buffer sized from a
decode-shaped dummy run and thus too small for a later prefill. Several defects
were latent because the reference text deployment runs with a single pipeline
stage, so the PP$>$1 code paths had never been exercised.

\paragraph{Root cause.} The vendor plugin is a young, fast-moving
reimplementation whose tested configurations are a subset of what upstream vLLM
supports. Feature combinations that are routine on CUDA (pipeline parallelism
$\times$ multimodal $\times$ MoE with a sparse-attention variant) fall outside the
tested envelope, and coverage gaps surface as hard crashes rather than graceful
fallbacks.

\subsection{Fragile multi-axis parallelism}
\label{sec:parallelism}
\paragraph{Symptom.} Splitting a single replica across 16 devices is constrained
and fragile. Two throughput-oriented parallel features had to be disabled to
obtain correct results, and the tensor-parallel degree is hard-capped. We use the
standard tensor/pipeline/expert-parallel decomposition of~\cite{megatron,gpipe,gshard}.

\paragraph{Evidence.}
\begin{itemize}
  \item \textbf{TP cap.} An attention head-grouping factor of~8 caps tensor
        parallelism at \code{TP\,=\,8}. To use all 16 devices for one replica a
        second axis is mandatory --- pipeline parallelism (\code{PP\,=\,2}) for the
        \code{bf16} multimodal model (Figure~\ref{fig:memory}), or data
        parallelism (\code{DP\,=\,2}) for the quantized LLM.
  \item \textbf{Sequence parallelism / FlashComm.} Enabling sequence parallelism
        (via the FlashComm path) shards hidden states by the TP degree, but the
        multimodal forward injects image embeddings through full-size
        \code{inputs\_embeds}; the sharded and full-size tensors then disagree
        (\code{expanded size (N) must match existing size (N/tp)}). We disabled it
        (Listing~\ref{lst:env}).
  \item \textbf{Fused MC2 expert routing.} The fused MoE communication kernel
        reads token ids from the forward context, which are absent during the
        multimodal memory-profiling dummy run, so it had to be disabled as well.
  \item \textbf{Pipeline parallelism correctness.} Beyond the crashes in
        \S\ref{sec:maturity}, a heterogeneous-computing head dimension carried on
        the hidden-state and residual streams between layers was being collapsed
        on \emph{every} rank and re-expanded on the next, making the per-head
        states identical mid-network and silently corrupting output. The fix
        forwards the full multi-head stream across the pipeline boundary and
        collapses it only on the last stage --- a defect that only manifests with
        \code{PP\,>\,1}.
\end{itemize}

\begin{lstlisting}[caption={Correctness-over-throughput parallel toggles for the multimodal path.},label={lst:env},language=bash]
# Fused MC2 expert routing needs the forward-context token ids, which are
# None during multimodal memory profiling (inputs_embeds-only dummy run).
export VLLM_ASCEND_ENABLE_FUSED_MC2=0
# FlashComm1 turns on sequence parallelism. SP shards hidden_states by
# tp_size, but the multimodal forward injects image embeddings via full-size
# inputs_embeds, so the sharded context and full hidden_states disagree.
export VLLM_ASCEND_ENABLE_FLASHCOMM1=0
\end{lstlisting}

\paragraph{Root cause.} Parallelism features are implemented and validated largely
for the single-modality, single-pipeline-stage case. Their interaction with
multimodal input injection, hash-routed MoE, and the sparse-attention variant is
thin, so each additional parallel axis multiplies the number of untested code
paths. The safe posture is to minimize parallel features and prefer correctness
over the associated throughput gains.

\subsection{Kernel-level numerical and stability faults}
\label{sec:kernel}
\paragraph{Symptom.} Under sustained load the service intermittently aborted with
low-level device execution faults --- both ``aicore'' and ``vector core''
exceptions --- that killed the worker process and required a restart.

\paragraph{Evidence.} The faults surfaced as CANN inner errors with runtime result
codes \code{507015} (aicore) and \code{507035} (vector core), and an extended
message pointing at an MTE address violation (``\code{The DDR address of the MTE
instruction is out of range}''). Representative sanitized excerpts are shown in
Listings~\ref{lst:vec} and~\ref{lst:aic}.

\begin{lstlisting}[caption={Vector-core execution fault (runtime result 507035).},label={lst:vec}]
RuntimeError: npuSynchronizeDevice: NPU function error:
    AclrtSynchronizeDeviceWithTimeout, error code is 507035
[Error]: The vector core execution is abnormal.
EZ9999: Inner Error!
    there is an exception of aivec error, core id is 21,
    error code = 0x800000
    errorStr: The DDR address of the MTE instruction is out of range.
    Kernel task happen error, retCode=0x31, [vector core exception].
    wait event to be complete failed, runtime result = 507035
\end{lstlisting}

\begin{lstlisting}[caption={Aicore execution fault (runtime result 507015).},label={lst:aic}]
RuntimeError: npuSynchronizeDevice: NPU function error:
    AclrtSynchronizeDeviceWithTimeout, error code is 507015
[Error]: The aicore execution is abnormal.
EZ9999: Inner Error!
    there is an exception of fftsplus aivector error, core id is 9,
    error code = 0x800000
    errorStr: The DDR address of the MTE instruction is out of range.
    Kernel task happen error, retCode=0x26, [aicore exception].
    wait for compute device to finish failed, runtime result = 507015
\end{lstlisting}

\paragraph{Root cause.} These are kernel-internal memory-addressing faults raised
by the device, not application-level exceptions. Because they originate inside
compiled operators and are reported only as opaque codes and register dumps, they
are essentially undebuggable from the serving layer. Their intermittency (tied to
load and specific shapes) points to edge cases in the operator library rather than
a deterministic bug we could isolate. Operationally they must be treated as
expected faults to be detected and recovered from (\S\ref{sec:reliability}).

\subsection{Immature graph compilation}
\label{sec:graph}
\paragraph{Symptom.} The high-performance graph-execution path is either
unavailable or carries explicit stability warnings, so we frequently fell back to
eager execution.

\paragraph{Evidence.} On the multimodal model we ran with \code{enforce-eager} (no
graph capture) to avoid capture-time failures. On the quantized LLM we used the
full-decode ACL-graph mode, for which the engine emits an explicit
experimental-stage warning that capturing too many batch sizes can cause
out-of-memory errors or inference hangs (Listing~\ref{lst:fullgraph}). The engine
also logs that TorchInductor is not supported on this platform, so only ACL Graph
mode is available and there is no JIT kernel fusion in the decode path.

\begin{lstlisting}[caption={Full-graph capture carries an explicit experimental/OOM/hang warning.},label={lst:fullgraph}]
WARNING platform.py] FULL_DECODE_ONLY compilation enabled on NPU.
    use_inductor not supported - using only ACL Graph mode
WARNING platform.py]
  * WARNING: You have enabled the *full graph* feature.
  * This is an early experimental stage and may involve various unknown issues.
  * A known problem is that capturing too many batch sizes can lead to OOM
  * (Out of Memory) errors or inference hangs. If you encounter such issues,
  * consider reducing gpu_memory_utilization or manually specifying a smaller
  * batch size for graph capture.
\end{lstlisting}

\paragraph{Root cause.} Graph capture and kernel compilation are less mature than
their CUDA counterparts. Without an Inductor-equivalent JIT, uncovered shapes and
operators fall back to eager mode, and the experimental full-decode capture trades
stability for throughput. The practical consequence is a hard choice between the
robust-but-slow eager path and the fast-but-fragile capture path.

\subsection{Advanced-feature gaps}
\label{sec:advanced}
\paragraph{Symptom.} Advanced inference features that are stable on CUDA are newer
and less reliable here, and enabling them can destabilize serving.

\paragraph{Evidence.} We configured Multi-Token Prediction (MTP) speculative
decoding~\cite{speculative,medusa,eagle} with a single speculative token on the
quantized LLM. In this stack the feature is deprecation-warned and renamed across
versions, and speculative quantization settings are silently overridden by the
platform. Combined with the kernel faults of \S\ref{sec:kernel}, the safest
configuration disabled speculative decoding entirely, trading its throughput
benefit for reliability.

\paragraph{Root cause.} Speculative decoding requires tight coordination between a
draft path, the main model, and the scheduler; on a stack where each piece is
itself young, the composed feature is the least stable of all. Feature
availability also churns between plugin versions, so a working recipe is
version-locked.

\subsection{Performance and scalability ceilings}
\label{sec:perf}
\paragraph{Symptom.} Effective throughput saturates at a surprisingly low request
concurrency, and time-to-first-serve is long.

\begin{figure}[t]
\centering
\begin{tikzpicture}
\begin{axis}[
    width=0.78\linewidth, height=5.6cm,
    xlabel={Request concurrency (in-flight streams)},
    ylabel={Aggregate throughput (req/h)},
    xmin=0.5, xmax=6.5, ymin=0, ymax=3200,
    xtick={1,2,4,6},
    axis y line*=left,
    ylabel near ticks,
    legend style={at={(0.03,0.97)}, anchor=north west, font=\footnotesize, draw=none, fill=none},
]
\addplot+[mark=*, thick, color=accent] coordinates {
    (1,900) (2,1800) (4,2800) (6,1080)
};
\addlegendentry{Aggregate throughput}
\end{axis}
\begin{axis}[
    width=0.78\linewidth, height=5.6cm,
    xmin=0.5, xmax=6.5, ymin=0, ymax=60,
    xtick=\empty,
    axis y line*=right,
    ylabel={Waiting queue depth},
    ylabel near ticks,
    legend style={at={(0.03,0.80)}, anchor=north west, font=\footnotesize, draw=none, fill=none},
]
\addplot+[mark=square*, thick, dashed, color=accentb] coordinates {
    (1,0) (2,0) (4,1) (6,55)
};
\addlegendentry{Waiting queue}
\end{axis}
\end{tikzpicture}
\caption{Aggregate serving throughput (left axis, solid) and steady-state waiting
queue depth (right axis, dashed) versus request concurrency for the quantized-LLM
service. Throughput is non-monotonic: it peaks near a concurrency of four and then
collapses as the queue saturates beyond the ``sweet spot''.}
\label{fig:throughput}
\end{figure}
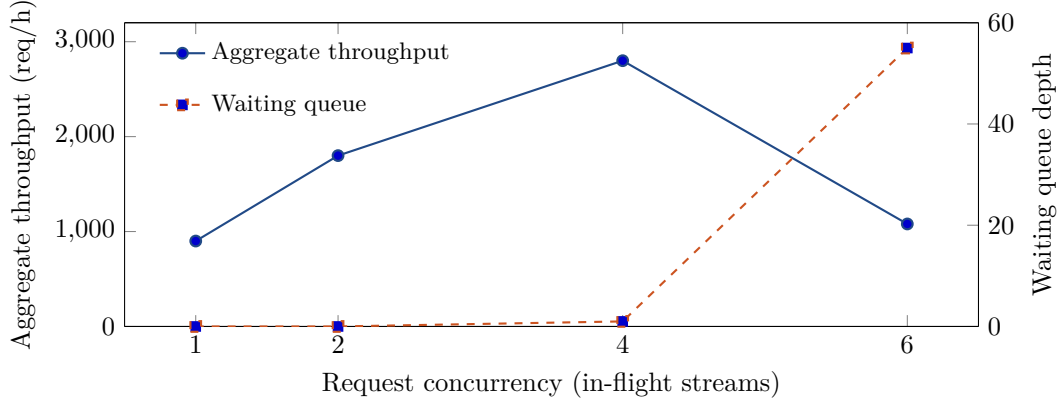

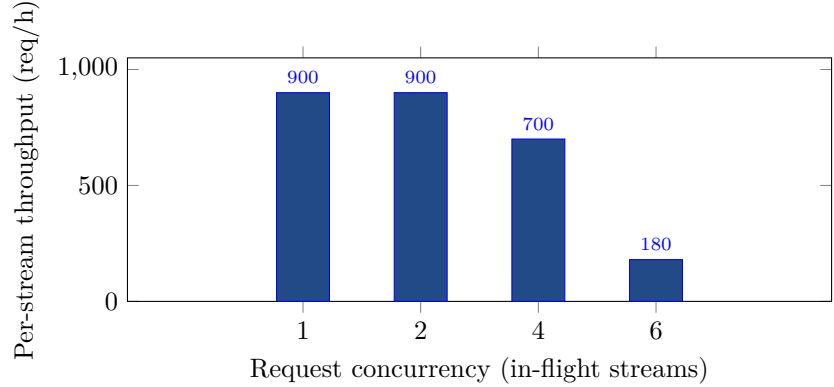
\begin{figure}[t]
\centering
\begin{tikzpicture}
\begin{axis}[
    width=0.66\linewidth, height=4.8cm,
    ybar,
    bar width=0.7cm,
    xlabel={Request concurrency (in-flight streams)},
    ylabel={Per-stream throughput (req/h)},
    xmin=0.3, xmax=4.7,
    ymin=0, ymax=1050,
    xtick={1,2,3,4},
    xticklabels={1,2,4,6},
    nodes near coords, nodes near coords style={font=\scriptsize},
    enlarge x limits=0.18,
]
\addplot+[fill=accent] coordinates {(1,900) (2,900) (3,700) (4,180)};
\end{axis}
\end{tikzpicture}
\caption{Per-stream throughput versus concurrency. Each additional stream past the
sweet spot receives a smaller share; at a concurrency of six, per-stream
throughput has dropped roughly five-fold, indicating that oversubscription
degrades rather than improves service.}
\label{fig:perworker}
\end{figure}

\paragraph{Evidence.} For the quantized-LLM service we found a concurrency ``sweet
spot'' of about four in-flight request streams. Below it the device was
underutilized; at that level the engine reported a near-zero waiting queue; at six
or more the queue saturated (dozens of requests waiting), parse/timeout failures
rose, and \emph{aggregate} throughput fell rather than rose.
Figure~\ref{fig:throughput} shows this non-monotonic behavior: aggregate
throughput peaks near a concurrency of four and then collapses as the waiting
queue explodes, while per-stream throughput
(Figure~\ref{fig:perworker}) degrades steadily under oversubscription.
Separately, cold-start took several minutes due to operator compilation and
weight/KV-cache quantization before the first request could be served.

\paragraph{Root cause.} The low concurrency ceiling reflects operator-library and
scheduler behavior that does not extract as much batch-level parallelism as the
CUDA path for these models, so oversubscription degrades instead of helping. The
slow startup reflects ahead-of-time operator compilation and quantization work
that the CUDA path either avoids or amortizes differently. Both are stack
properties, not tuning mistakes, and must be planned around (fixed modest
concurrency; long readiness timeouts).

\subsection{Operational reliability and observability}
\label{sec:reliability}
\paragraph{Symptom.} Long-running services do not stay up on their own, and when
they fail the diagnostics are cryptic.

\paragraph{Evidence.} We ran the serving process under an external watchdog that
health-checks the OpenAI-compatible endpoint every 30 seconds and automatically
restarts the engine on failure --- without it, the kernel faults of
\S\ref{sec:kernel} would have caused unbounded outages. We also observed the
engine repeatedly logging that no shared-memory broadcast block was available for
60 seconds at a time (Listing~\ref{lst:shm}), a symptom of a process hanging or
doing long compilation/quantization work, with no finer-grained signal as to
which. Combined with the opaque device fault dumps, root-causing incidents from
logs alone was rarely possible.

\begin{lstlisting}[caption={Repeated shared-memory broadcast stall, indicating a hang or long-running compilation/quantization step.},label={lst:shm}]
INFO shm_broadcast.py] No available shared memory broadcast block
    found in 60 seconds. This typically happens when some processes are
    hanging or doing some time-consuming work (e.g. compilation,
    weight/kv cache quantization).
\end{lstlisting}

\paragraph{Root cause.} The stack surfaces failures as low-level device or IPC
messages rather than actionable, layered diagnostics, and it lacks the mature
resiliency features one expects from a production serving path. Reliability
therefore has to be supplied externally (watchdogs, restart policies, log scraping
for known fault signatures).

\subsection{Ecosystem fragmentation and portability tax}
\label{sec:ecosystem}
\paragraph{Symptom.} Getting and keeping a working deployment depends on a matrix
of vendor forks, environment variables, and startup patches that must be
rediscovered and pinned.

\paragraph{Evidence.} Behavior is governed by numerous vendor-specific environment
variables (for example a family of \code{VLLM\_ASCEND\_ENABLE\_*} feature toggles,
plus \code{ACL\_OP\_INIT\_MODE}, \code{HCCL\_BUFFSIZE}, and NPU allocator knobs),
and by model-/engine-specific source patches applied at container startup. Correct
operation depends on a specific engine version, a specific patch set, and a
specific env-var configuration; changing any one can silently alter correctness or
stability.

\paragraph{Root cause.} Because the vendor path is a fork/plugin that trails and
diverges from upstream, portability from a CUDA deployment is not free: it incurs a
recurring ``tax'' of re-patching, re-tuning, and re-validating on every version
bump. The small community and thin public documentation make this tax higher than
for the mainstream ecosystem.

\section{Discussion}
\label{sec:discussion}
The eight categories share a small number of cross-cutting root causes.
\emph{First}, an immature compiler and operator library is upstream of the kernel
faults (\S\ref{sec:kernel}), the graph-compilation limits (\S\ref{sec:graph}), and
the low concurrency ceiling and slow startup (\S\ref{sec:perf}). \emph{Second},
thin coverage of feature \emph{combinations} --- as opposed to individual features
--- explains the parallelism fragility (\S\ref{sec:parallelism}) and most of the
startup patches (\S\ref{sec:maturity}): each feature may work in isolation, but
their cross-product is untested. \emph{Third}, missing production-grade reliability
and observability (\S\ref{sec:reliability}) turns the inevitable faults into
outages unless external scaffolding is added. \emph{Fourth}, ecosystem
fragmentation (\S\ref{sec:ecosystem}) amplifies all of the above by making every
fix version-specific and hard to share.

A recurring and dangerous pattern is that startup can look healthy while inference
is broken: memory profiling and warmup exercise different code paths than real
requests, so several defects only appeared on the first real prefill or under
sustained load. Teams should therefore validate with realistic traffic, not just a
successful launch.

We stress that none of this means the platform is unusable --- both workloads were
ultimately served correctly. It means the \emph{total cost of ownership} is
dominated by engineering effort (patching, tuning, operating) rather than by
acquiring the hardware, and that budgeting for that effort is the single most
important planning decision.

\subsection{Threats to validity}
\label{sec:threats}
Our study is observational and has clear limits. It covers a \emph{single vendor}
(Huawei Ascend), a \emph{single engine line} (vLLM-Ascend 0.13.x), and a
\emph{single node} of 16 devices; other accelerators, engine versions, or cluster
scales may behave differently, and several defects we hit are likely to be fixed
in later releases. The workloads are \emph{two specific architectures} (a
W8A8 MoE LLM and an MoE-decoder VLM) that deliberately stress the hardest paths, so
the incidence of problems is not representative of simpler dense models. Our
performance measurements are coarse operational observations rather than
controlled benchmarks, and absolute numbers depend on request mix and generation
lengths; we therefore report them only to illustrate \emph{qualitative} behavior
(e.g., the existence of a concurrency sweet spot) rather than as precise figures.
The benchmark scores in Figure~\ref{fig:bench} are likewise used only to establish
correct operation, not as a model-quality claim. We believe the \emph{classes} of
limitation and the cross-cutting root causes generalize even where specific codes
or counts do not.

\section{Recommendations and General Strategies}
\label{sec:mitigations}
We separate concrete tactics for the specific stack from general strategies that
apply to adopting \emph{any} alternative accelerator.

\subsection{Tactics for this stack}
\begin{enumerate}
  \item \textbf{Adopt a correctness-first configuration.} Disable fragile
        throughput features by default (sequence parallelism/FlashComm, fused MoE
        communication, speculative decoding) and re-enable them only after
        workload-specific validation. Prefer \code{enforce-eager} unless graph
        capture is proven stable for your shapes.
  \item \textbf{Minimize parallel axes.} Respect the tensor-parallel cap and add
        only the one extra axis (PP or DP) you actually need; every extra parallel
        feature multiplies untested code paths.
  \item \textbf{Wrap the service in a watchdog.} Assume kernel faults will occur;
        health-check the endpoint and auto-restart. Scrape logs for known fault
        signatures (e.g., runtime results \code{507015}/\code{507035}, MTE
        out-of-range) to drive alerting.
  \item \textbf{Tune to a concurrency sweet spot.} Measure the point where the
        waiting queue stays near zero (Figure~\ref{fig:throughput}) and fix
        concurrency there rather than oversubscribing.
  \item \textbf{Pin everything and keep a patch set.} Lock the engine/plugin
        version, keep source patches idempotent and applied at startup, and
        document the full environment-variable matrix so the deployment is
        reproducible across restarts and hosts.
\end{enumerate}

\subsection{General strategies for adopting alternative accelerators}
These strategies are vendor-agnostic and, in our experience, are what actually
bound the cost and risk of a migration.
\begin{enumerate}
  \item \textbf{Differential testing against a reference.} Treat a trusted
        implementation (e.g., the CUDA or CPU path, or the model's native
        framework) as a correctness oracle and diff intermediate tensors and final
        outputs layer by layer. Many of our hardest bugs were silent numerical
        corruptions, not crashes; only differential comparison catches these.
  \item \textbf{Continuous integration on the target stack.} Run the real
        prefill/decode paths (not just startup) on target hardware in CI for every
        engine/plugin bump, so coverage gaps and regressions are caught before
        production.
  \item \textbf{A thin portability layer.} Isolate engine- and vendor-specific
        configuration (env vars, flags, patches) behind a small abstraction so the
        rest of the serving system is portable and the ``portability tax'' is paid
        in one place.
  \item \textbf{Canary and shadow deployment.} Roll out new versions/patch sets to
        a canary that mirrors production traffic, and shadow real requests to
        detect quality or stability regressions before full rollout.
  \item \textbf{A feature-flag compatibility matrix.} Maintain an explicit,
        tested matrix of which feature combinations (parallelism $\times$ modality
        $\times$ MoE $\times$ quantization $\times$ graph mode) are known-good on
        each version, and default to the smallest working set.
  \item \textbf{Capacity planning around measured limits.} Size fleets from the
        measured concurrency sweet spot and cold-start time
        (Figures~\ref{fig:throughput}--\ref{fig:perworker}) rather than from
        theoretical peak FLOPs or memory.
  \item \textbf{Fault-tolerant serving by default.} Assume device faults; combine
        health-checked auto-restart, request retries/idempotency, and
        fault-signature alerting so that isolated kernel faults degrade gracefully
        instead of causing outages.
  \item \textbf{Kernel-level fallback policies.} Prefer the most robust execution
        mode (eager) as a safe default and adopt graph capture or fused kernels
        selectively, with automatic fallback when capture fails.
  \item \textbf{Upstream and collaborate.} Contribute fixes and reproducers back to
        the vendor plugin and engage the community; on a thin ecosystem, shared
        fixes reduce everyone's recurring cost and shorten the divergence from
        upstream.
\end{enumerate}

\paragraph{Asks for the ecosystem.} Graceful fallbacks (or clear errors) instead
of hard crashes for uncovered feature combinations; layered, actionable
diagnostics that map device faults to the offending operator/layer; broader
testing of feature cross-products (parallelism $\times$ multimodal $\times$ MoE
$\times$ sparse attention); and tighter, better-documented version/feature
compatibility matrices would each remove a large share of the cost documented
here.

\section{Conclusion}
\label{sec:conclusion}
We reported a field study of deploying a large W8A8 MoE language model and a large
multimodal vision--language model for inference on a 16-card Huawei Ascend~910
system using CANN and vLLM-Ascend. Both were eventually served correctly, but only
after twelve source patches, the deliberate disabling of several performance
features to preserve correctness, and external operational scaffolding to absorb
recurring low-level device faults. The limitations we encountered --- stack and
operator/feature-coverage immaturity, fragile multi-axis parallelism, kernel-level
aicore/vector-core faults, immature graph compilation, unstable advanced features,
low performance/scalability ceilings, weak reliability/observability, and ecosystem
fragmentation --- are platform properties rather than workload quirks. For teams
considering this class of accelerator, the hardware is capable, but the engineering
cost of reaching and maintaining correctness is the dominant consideration and
should be planned for explicitly. The general strategies in
Section~\ref{sec:mitigations} --- differential testing, CI on the target stack,
portability layers, canary/shadow rollout, capacity planning around measured
limits, and fault-tolerant serving --- are, in our experience, what make that cost
manageable.

\bibliographystyle{plainnat}
\bibliography{references}

\end{document}